\documentclass[aps,prl,reprint,groupedaddress,nofootinbib,longbibliography,showpacs]{revtex4-2}
\usepackage{graphicx}
\usepackage{latexsym,amssymb,amsmath}
\usepackage{bm}
\usepackage{braket}
\usepackage[caption=false]{subfig}
\usepackage{xcolor}
\usepackage{mathrsfs}
\usepackage[colorlinks=true,
urlcolor=blue,
linkcolor=blue,
citecolor=blue
]{hyperref}

\begin{document}
\title{High-order topological quantum optics in ultracold atomic metasurfaces}
\author{B. X. Wang}
\affiliation{Institute of Engineering Thermophysics, School of Mechanical Engineering, Shanghai Jiao Tong University, Shanghai 200240, China}
\author{C. Y. Zhao}
\email{changying.zhao@sjtu.edu.cn}
\affiliation{Institute of Engineering Thermophysics, School of Mechanical Engineering, Shanghai Jiao Tong University, Shanghai 200240, China}

\date{\today}

\begin{abstract}
Ultracold atom arrays in optical lattices emerge as an excellent playground for the integration of topological photonics and quantum optics. Here, we study high-order topological quantum optics in an ultracold atom metasurface intended to mimic the two-dimensional Su-Schrieffer-Heeger model. We find the existence of long-range interactions beyond nearest-neighbor ones leads to isolated corner states in the band gap, and show a corner atom can be addressed by a laser drive far away from it via these nontrivial states. We demonstrate the Purcell factor can be used as a powerful tool to examine the existence of topological edge and corner states. We predict topological edge states can mediate strong coherent interactions between two remote impurity quantum emitters while suppressing dissipative losses thanks to the higher-order topology, generating robust and long-lived quantum entanglement, without the need for additional photonic structures.
\end{abstract}
\maketitle
Topological photonics, distinctive in the topologically protected unidirectional edge states immune against disorder and imperfections \cite{khanikaevNaturemat2013,luNPhoton2014,khanikaevNPhoton2017}, opens new prospects of realizing resilient photonic devices, such as unidirectional waveguides \cite{poliNComms2015}, optical isolators \cite{el-GanainyOL2015,KarkiPRApplied2019}, topological lasers \cite{partoPRL2018,zengNature2020,shaoNaturenano2020} and topological sensors \cite{wangPRM2020}. The application of topological photonics in the quantum domain enables surprisingly robust and efficient avenues for quantum technologies. Topological photonic states can mediate strong coupling and entanglement between distant quantum emitters (QEs) \cite{wangPRApp2020}, induce or protect multiphoton quantum interferences \cite{tambascoSciAdv2018,blancoredondoScience2018}, quantum correlations \cite{wangOptica2019} and quantum coherence \cite{niePRL2020}, and lead to substantial non-reciprocal photon scattering \cite{niePRApp2020}. These new advances constitute an emerging multidisciplinary field dubbed topological quantum optics \cite{wangPRA2021,barikScience2018,doyeuxPRL2017,mivehvarPRL2017,panPRL2015}. 

Previous works mainly rely on the integration of photonic structures and QEs, while deterministic interfacing them has been technically difficult, especially in a scalable manner \cite{changRMP2018,solntsevNaturephoton2021}. Alternatively, the intrinsic quantum nature of atomic arrays, that is, a single atom cannot be excited twice, permits an ideal playground for this emerging field. Recent progress in producing ordered cold atomic arrays \cite{roatiNature2008,schreiberScience2015,bordiaNaturephys2017,endresScience2016,lahayeScience2016}, thanks to current laser cooling and trapping technologies \cite{blochNaturephys2005,lesterPRL2015}, makes it possible to achieve an extremely efficient manipulation of photon-atom interactions by using only a small number of atoms, since significant spatial interferences of emitted photons arising from specific ordered arrangement can mediate a variety of exotic collective and cooperative excitations \cite{ballantinePRResearch2020,massonPRL2020,yelinPRL2017,ruiNature2021,ryanPRL2020,glicensteinPRL2020}. These arrays are naturally suitable for parallel quantum operations \cite{bekensteinNaturephys2020}, high-fidelity quantum information storage \cite{facchinettiPRL2016}, entanglement generation \cite{guimondPRL2019}, photon-photon correlations \cite{morenocardoner2021quantum} as well as quantum optical metamaterials or metasurfaces \cite{ballantinePRL2020,alaeePRL2020,wangOE2017}. 
Topological quantum optical states have also been demonstrated by several recent works including the integer quantum Hall effect in 2D atomic arrays \cite{bettlesPRA2017,yelinPRL20172,perczelPRA2017}, and long-lived topological states in 1D dimerized atomic chains \cite{zhangCommPhys2019,zhangPRL2019,zhangPRL2020,wang2018topological}. Crucially, the intrinsic nonlinearity of photon-atom interaction can be exploited to induce strong photon-photon interactions, offering a platform for achieving many-body topological states of photons such as those in the fractional quantum Hall effect \cite{umucalilarPRL2012,carusottoRMP2013,maghrebiPRA2015,roushanNaturephys2017,perczelPRL2020}. 

Recent discovery of high-order topological insulators (HOTIs) have led to novel aspects of topological photonics. In general, a $m$-th order topological insulator (TI) in $n$D can have $n$D gapped bulk states and $(n-1)$D, $(n-2)$D, ..., $(n-m-1)$D gapped edge states while showing $(n-m)$D gapless corner states. Such multidimensional topological physics lead to very interesting physics beyond conventional bulk-edge correspondence, also offering emerging opportunities for the trapping, waveguiding and lasing of photons \cite{schindlerSciAdv2017,benalcazarScience2017,benalcazarPRB2017,serragarciaNature2018,mittalNaturephoton2019,xueNaturemater2019,niNaturemater2019,elhassanNaturephoton2019,petersonNature2018,liNaturephton2019,petersonScience2020,cerjanPRL2020}. However, a quantum optical counterpart in the cold atomic platform remains to be unexplored.

\begin{figure}[htbp]
	\centering
	\includegraphics[width=1\linewidth]{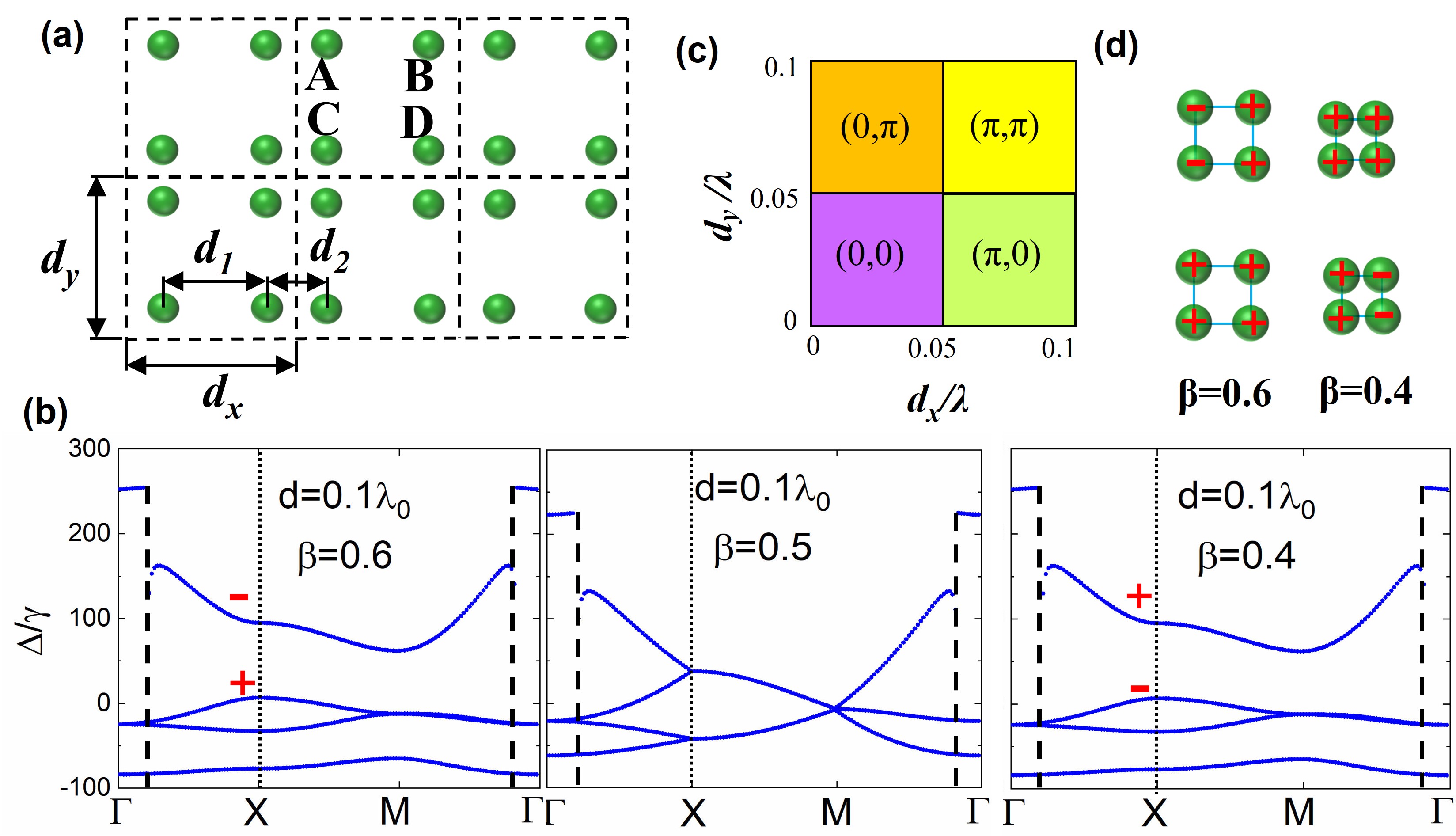}
	
	\caption{Bulk properties and topology. (a) The cold atom metasurface mimicking the 2D SSH model. (b) Band structures for different dimerization parameters. $+$ ($-$) implies the even (odd) parity of the band.  (c) 2D Zak phase for different system parameters $d_x$ and $d_y$ for the second band gap ($N_\mathrm{occ}=3$). (d) Polarization directions (along $+z$ or $-z$) of the atomic eigenstate at $X$ point in the third (lower) and fourth (higher) bands for $\beta=0.6$ and $0.4$. $+$ and $-$ correspond to the sign of $p_{\sigma,\mathbf{k}}$ and thus the parities of eigenwavefunctions can be determined (c.f. Fig. \ref{fig1band}(b)). The periodicity is $d=0.1\lambda_0$.}
	\label{fig1band}
\end{figure}

Here, we consider high-order topological quantum optics in a two-dimensional ultracold atom metasurface, which is intended to mimic the 2D Su-Schrieffer-Heeger (SSH) model but have more profound implications as discussed below.  
The schematic of atomic metasurface is presented in Fig. \ref{fig1band}(a), where atoms are assumed to be well trapped in a perfect Mott insulator state in an optical lattice \cite{blochNaturephys2005}. It is a square lattice followed by a dimerization procedure in each dimension by introducing inequivalent intercell and intracell spacings, i.e., $d_1\neq d_2$, resulting four sublattices denoted by $A$, $B$, $C$ and $D$ respectively. The overall period is $d_x=d_y=d=d_1+d_2$ with the definition of dimerization parameter being $\beta=d_1/(d_1+d_2)$. Such dimerization leads to different ``hopping" amplitudes of photons along different directions like those in conventional 2D SSH model \cite{liuPRL2017}.  The two-level atom is assumed to have three degenerate excited states denoted by $|e_{\alpha}\rangle$ polarized along different directions, where $\alpha=x,y,z$ stands for Cartesian coordinates, with a ground state denoted by $|g\rangle$. By applying the single and weak excitation approximations, we can work in the subspace spanned by the ground states and single excited states of the atoms \cite{kaiserJMO2011,kaiserFP2012,guerinPRL2016}. By adiabatically eliminating the photonic degrees of freedom, the effective Hamiltonian is given by \cite{antezzaPRL2009,antezzaPRA2009,kaiserJMO2011,kaiserFP2012,guerinPRL2016} 
\begin{equation}\label{Hamiltonian}
\begin{split}
H&=\hbar\sum_{i=1}^N\sum_{\alpha=x,y,z}(\omega_0-i\frac{\gamma}{2})\ket{e_{i,\alpha}}\bra{e_{i,\beta}}\\&+\frac{3\pi\hbar\gamma c}{\omega_0}\sum_{i=1,i\neq j}\sum_{\alpha,\beta=x,y,z}G_{\alpha\beta}(\mathbf{r}_j,\mathbf{r}_i)\ket{e_{i,\alpha}}\bra{e_{j,\beta}},
\end{split}
\end{equation}
where $\hbar$ is the Planck's constant, $\omega_0$ is transition frequency of a single atom in free space with a radiative linewidth of $\gamma$, and $c$ is the speed of light in vacuum. $G_{\alpha\beta}(\mathbf{r}_j,\mathbf{r}_i)$ is the free-space dyadic Green's function describing the propagation of field emitting from the $i$-th atom to $j$-th atom, where $\mathbf{r}_j$ and $\mathbf{r}_i$ indicate their positions \cite{SM}.
In the following we consider out-of-plane modes in which all atoms are excited to $|e_z\rangle$ states, assuming in-plane polarizations can be shifted to other frequencies. By invoking Bloch theorem, we construct the eigenwavefunction with an in-plane wavevector $\mathbf{k}=k_x\hat{x}+k_y\hat{y}$ as a linear combination of the single-excited states as
$|\psi_{\mathbf{k}}\rangle=\sum_{n=-\infty}^{\infty}\exp(i\mathbf{k} \cdot\mathbf{R}_{n})[p_{A,\mathbf{k}}\ket{e_{nA,z}}+p_{B,\mathbf{k}}\ket{e_{nB,z}}+p_{C,\mathbf{k}}\ket{e_{nC,z}}+p_{D,\mathbf{k}}\ket{e_{nD,z}}]$,
where $n$ denotes the $n$-th unit cell, $|e_{n\sigma,z}\rangle$ stand for the single excited states of the $\sigma$-type atom with $\sigma=A,B,C,D$, and $p_{\sigma,\mathbf{k}}$ denotes expansion coefficients depending on $\sigma$ and $\mathbf{k}$. We solve non-Hermitian eigenstate problem $H|\psi_\mathbf{k}\rangle=\hbar E_{\mathbf{k}}|\psi_\mathbf{k}\rangle$ to obtain the complex eigenfrequency $E_\mathbf{k}$ of the Bloch eigenstate, described by $E_\mathbf{k}=\omega_\mathbf{k}-i\Gamma_\mathbf{k}/2$ with $\omega_\mathbf{k}$ denoting the angular frequency and $\Gamma_\mathbf{k}$ the radiative linewidth [see details in Ref. \cite{SM}]. 
We further denote $\Delta=\mathrm{Re}E_{\mathbf{k}}-\omega_0$ as the detuning of the eigenstate, and $\Gamma=-2\mathrm{Im}E_{\mathbf{k}}$ as the decay rate. Note this model takes fully retarded near-field and far-field dipole-dipole interactions into account beyond the nearest-neighbor approximation \cite{wangPRB2018b}.



Figure \ref{fig1band}(b) shows Bloch band structures of a subwavelength atomic metasurface with $d=0.1\lambda_0$ for $\beta=0.6,~0.5$ and $0.4$ where $\lambda_0=2\pi c/\omega_0$ is the transition wavelength. Four bands are observed, and near the light line due to the strong coupling with free-space radiation the fourth band shows discontinuities \cite{yelinPRL20172}. For $\beta=0.5$ the third and fourth bands are degenerate at high-symmetry points of $X$ and $M$ as a consequence of band folding. By introducing dimerization (namely $\beta\neq0.5$), band degeneracies are broken and gaps are opened, and a smaller period $d$ and larger dimerization degree $|\beta-0.5|$ can lead to a wider band gap due to enhanced dipole-dipole interactions at small interatomic spacings \cite{wang2018topological}. Although chiral symmetry is broken due to intercell couplings \cite{kimNaturecomms2020,SM}, the system still respects $C_{4v}$ crystalline symmetry, protecting band topology and leading to two topological phases \cite{kimNaturecomms2020}. 

The bulk band topology of the 2D SSH model, which stems from the quantization of dipole moments \cite{benalcazarScience2017}, can be characterized by the 2D Zak phase \cite{chenPRL2019,chenPRB2020,xiePRL2019}. The 2D Zak phase is associated with the bulk polarization $P_j$ in terms of $\theta_j=2\pi P_j$ [see details in Ref. \cite{SM}]. In Fig. \ref{fig1band}(c), one can clearly observe the Zak phase for the second band gap is quantized for a set of system parameters $d_x$ and $d_y$. Here we do not require $d_x$ and $d_y$ to be equal. For $d_x,~d_y>0.5d$, the Zak phase becomes $(\pi,\pi)$, which, according to bulk-edge correspondence, implies the emergence of edge states in both $x$ and $y$ boundaries. Furthermore, according to the edge-corner correspondence, this set of Zak phase also indicates the existence of corner states. The  topological corner charge can be determined by the edge polarizations and then the bulk polarizations as $Q_c=4P_xP_y$, which is understood as the number of corner modes for each corner \cite{xiePRB2018}. On the other hand, for $d_x,~d_y<0.5d$, no edge or corner states can be observed. For other two cases ($d_x<0.5d$, $d_y>0.5d$ or the reverse), there are only edge states in one dimension and no corner states can emerge \cite{liuPRB2018,xiePRB2018,kimPRB2020}.

\begin{figure}[htbp]
	\centering
	\includegraphics[width=1\linewidth]{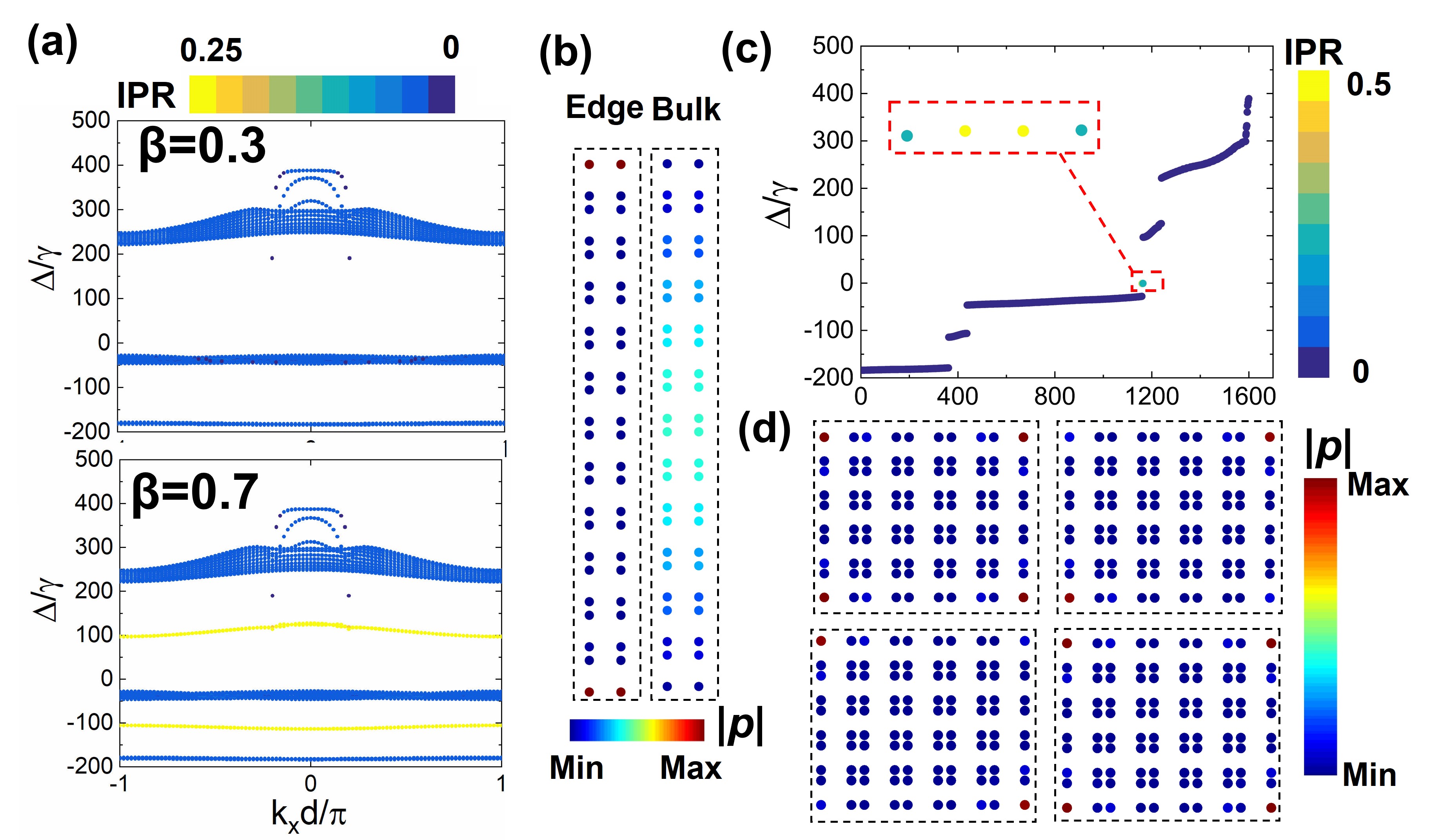}
	
	\caption{Edge and corner states. (a) Band structures under periodic boundary condition in $x$-axis with 12 unit cells in the $y$-direction for $\beta=0.3$ and 0.7. (b) Comparison between the edge and bulk states in the topological case. (c) Eigenstates of a fully open system. (d) Wavefunctions of four corner states.}
	\label{fig2edgecorner}
\end{figure}

Figure \ref{fig2edgecorner} shows verification results of above hierarchy of bulk-edge-corner correspondence \cite{liuPRB2018,xiePRB2018,kimPRB2020}. Consider a supercell with open edges, in which 12 unit cells aligned along the $y$-axis and periodic boundary condition is imposed along the $x$-direction [see details in Ref. \cite{SM}]. A comparison between band structures of topologically trivial and nontrivial cases is presented in Fig. \ref{fig2edgecorner}(a). Here the inverse participation ratio (IPR) of an eigenstate in the band structure is calculated, which is defined as $\mathrm{IPR}=\sum_{j=1}^{N}|p_j|^4/(\sum_{j=1}^{N}|p_j|^2)^2$ for a generalized wavefunction containing $N$ atoms $|\psi\rangle=\sum_{j=1}^{N}p_{j}|e_{j,z}\rangle$, where $p_{j}$ is the probability amplitude of $|e_{j,z}\rangle$. The IPR depicts the spatial confinement degree of eigenstates \cite{Skipetrov2014}. In Fig. \ref{fig2edgecorner}(a) for the topologically nontrivial case ($\beta=0.7$), topological edge states with relatively high IPRs ($\sim0.25$) appear in the first and second band gaps respectively, with absolute values of wavefunction coefficients at different lattice sites shown in Fig. \ref{fig2edgecorner}(b), compared with those of a randomly selected bulk state, exhibiting a strong edge localization behavior. The eigenstates of a topologically nontrivial $40\times40$ atomic lattice with fully open boundary conditions in both $x$- and $y$-directions are presented in Fig. \ref{fig2edgecorner}(c). Besides two bands of edge states residing in the first and second gaps, four corner states emerge near $\Delta/\gamma\sim-0.6$, in the gap between the third bulk and the second edge band. Their wavefunctions (here for readability a $10\times10$ lattice is presented) shown in Fig. \ref{fig2edgecorner}(d) confirm they are localized over the corners. By increasing $\beta$ from 0.6 to 0.7, we find the spectral position of topological corner states gradually move from within the bulk bands into isolated corner states. This phenomenon is not previously reported, which is due to the strong intercell dipole-dipole interactions. 
Without these interactions breaking chiral symmetry, corner states would be embedded in the continuum of bulk modes \cite{chenPRB2019,benalcazarPRB2020,liNJP2021}.

\begin{figure}[htbp]
	\centering
	\includegraphics[width=1\linewidth]{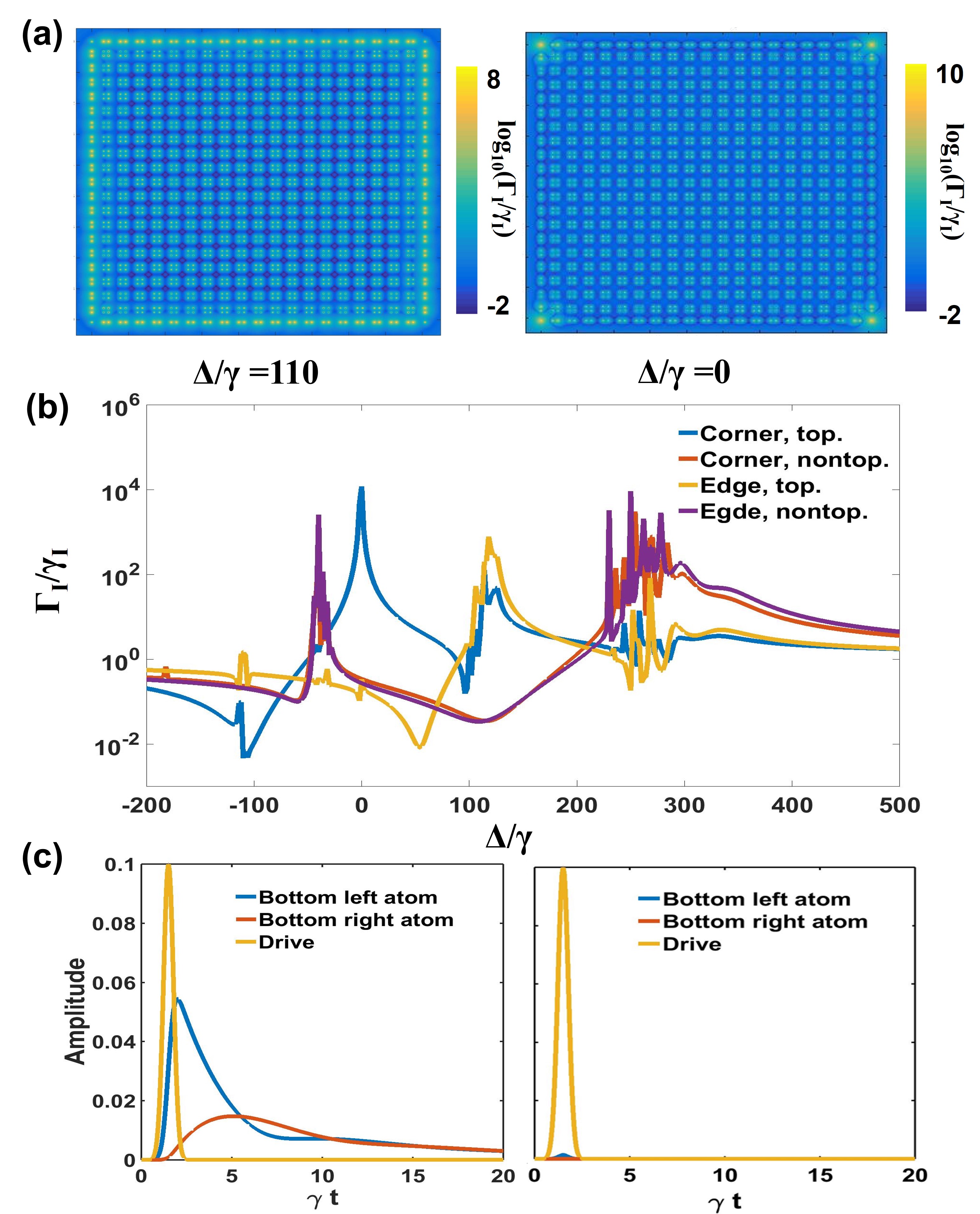}
	
	\caption{Corner states and Purcell factor. The Purcell factor of an impurity QE with a detuning of (a) $\Delta/\gamma=110$ and $\Delta/\gamma=0$ at different positions in the cold atomic metasurface. (b) The Purcell factor spectra for an impurity QE placed near the corner or the middle of the edge of topological and non-topological atomic metasurfaces. (c) The excitation dynamics of corner atoms. An indirect excitation of remote atoms mediated by topological corner states is found in the left panel. Comparison between the topologically trivial case is given in the right panel.}
	\label{fig3cornerLDOS}
\end{figure}

Next we show the existence of atomic metasurface can significantly affect the photonic environment by studying the Purcell factor for an impurity QE near it, which is the ratio between the actual decay rate $\Gamma_I$ and the free-space value $\gamma_I$. The result is obtained via the effective Green's function in the presence of cold atomic metasurface. This approach neglects non-Markovian effects which may lead to non-exponential decay of spontaneous emission, valid when the effective Green's function does not vary significantly in the range of impurity QE's linewidth, $\gamma_I\ll\gamma$ [see Ref. \cite{SM}]. In Fig. \ref{fig3cornerLDOS}(a), two situations are considered where transition frequency $\omega_I$ of the impurity QE is $\Delta/\gamma=110$ ($\Delta=\omega_I-\omega_0$, near the frequency of the edge state) and $\Delta/\gamma=0$ (near the corner states) respectively. Clear signatures of edge and corner states are observed. Moreover, the spectra of Purcell factor for an impurity QE placed near the corner or the middle of the edge of topological and non-topological atomic metasurfaces are given in Fig. \ref{fig3cornerLDOS}(b). In the topologically nontrivial case, a significant peak is observed near the frequency of corner states for the impurity QE near the corner, which disappears for the impurity QE near the edge. In the meanwhile, near the frequency of edge states ($\Delta\sim\pm110\gamma$), the impurity QE near the edge exhibits more significant Purcell factor. On the contrary, in the topologically trivial case, the spectra of impurity QEs near the edge and corner almost overlap with each other, exhibiting enhancements only near frequencies of bulk bands. Therefore by mapping of Purcell factor one can clearly demonstrate the existence of edge and corner states, offering a powerful tool for identifying higher-order topology \cite{cerjanPRL2020}.

We then demonstrate a unique property of topological corner states, where a corner atom can be addressed by an external laser driving another corner atom remotely. A Gaussian-shape source field, $\Omega(t)=\Omega\exp{(-[t-1.5\gamma^{-1}]^2/[0.15\gamma^{-2}])}$ for $t<1.5\gamma^{-1}$, is applied to excite the bottom left corner atom, where the driving frequency is $\Delta=0$ to excite the corner state. The Rabi frequency of the source field is chosen to be small enough as $\Omega=0.2\gamma$ so that the excitation of the system follows mainly on the collective decay from the system rather than the Rabi oscillation \cite{zhangCommPhys2019,yelinPRL20172}. 
It is obvious that only corner atoms are excited, which means a corner atom can be addressed by a laser drive far away from it due to topological corner states [left panel of Fig. \ref{fig3cornerLDOS}(c)]. In contrast, for the topologically trivial atomic lattice, this remote excitation phenomenon does not exist [right panel of Fig. \ref{fig3cornerLDOS}(c)]. The time evolution of the wavefunction is also presented in Fig. S2 \cite{SM}.

\begin{figure}[htbp]
	\centering
	\includegraphics[width=1\linewidth]{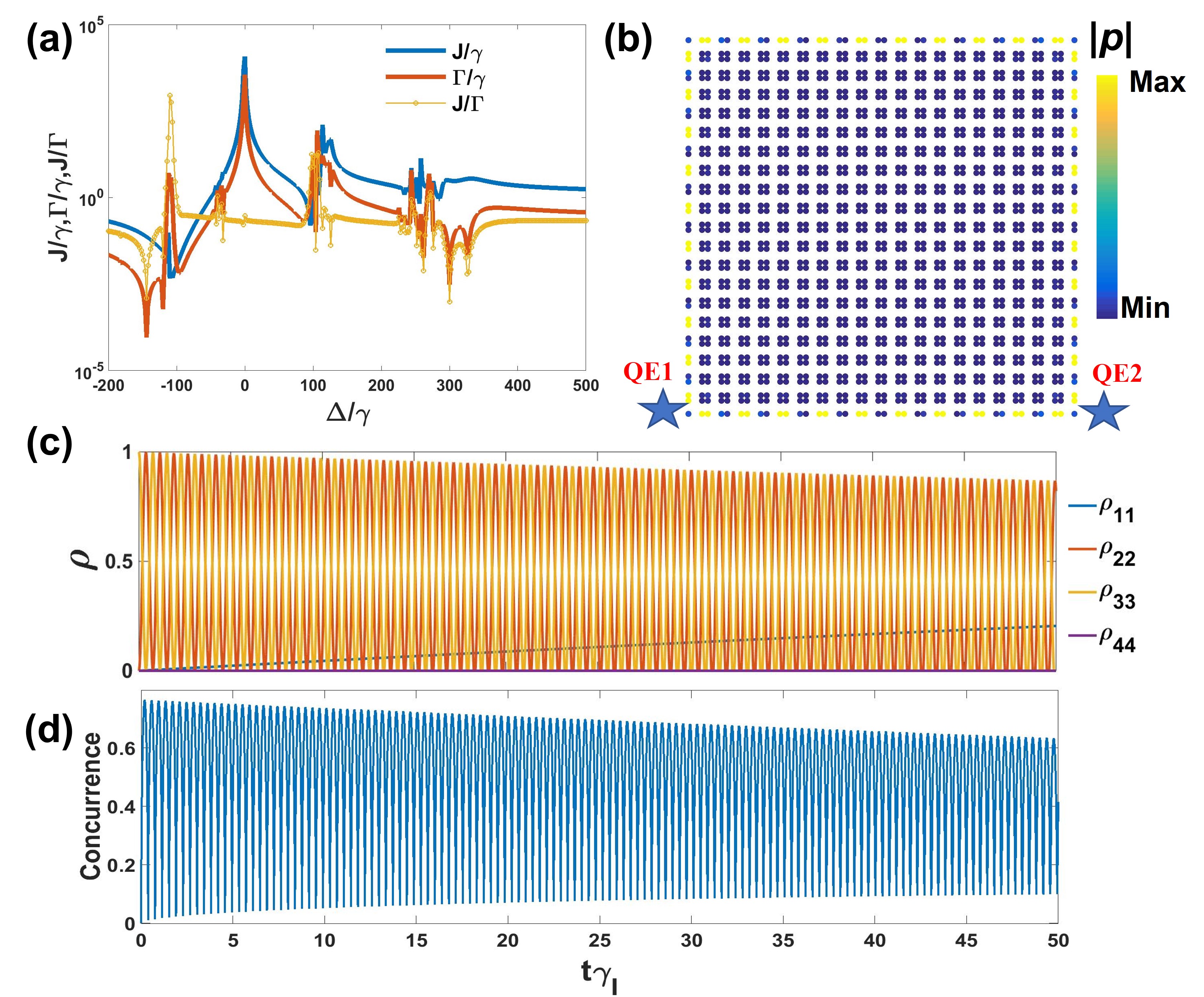}
	
	\caption{Entanglement generation from the cold atomic metasurface.  (a) The absolute value of coherent interaction $J_{12}$, decay rate $\Gamma_I$ and their ratio $J_{12}/\Gamma$ for different QE transition frequencies represented by detuning $\Delta$. (b) Wavefunction of the topological edge state at $\Delta/\gamma=-110$. The stars indicate positions of two QEs. (c) Time evolution of density matrix elements. (d) Time evolution of concurrence.}
	\label{edgeconcurrence}
\end{figure}

Finally, we demonstrate high-order topology in cold atomic metasurfaces can mediate strong coherent interactions for impurity QEs, leading to significant quantum entanglement generation. Consider a lattice composed of $40\times 40$ atoms with $d=0.1\lambda_0$ and $\beta=0.7$. One impurity QE is located away from the bottom left corner atom with $(-0.3d,0)$ and the other identical QE is displaced from the bottom right corner atom with $(0.3d,0)$. Assume $\gamma_I\ll\gamma$ again to make the characteristic time of two-QE dynamics is much longer than that of cold atomic metasurface, validating the Markovian approximation \cite{asenjogarciaPRA2017}. We calculate the coherent interaction $J_{12}=(3\pi\gamma_I/k)\mathrm{Re}\left[G_{zz}(\mathbf{r}_1,\mathbf{r}_2)\right]$ between the two QEs as well as the spontaneous decay rate $\Gamma_{11}=\Gamma_{22}=\Gamma_{I}$ of themselves as a function of the detuning $\Delta=\omega_\mathrm{I}-\omega_0$, where $G_{zz}(\mathbf{r}_1,\mathbf{r}_2)$ is the effective Green's function in the vacuum modified by the presence of cold atomic metasurface. The spectra of $J_{12}$ and $\Gamma_\mathrm{I}$ along with their ratio are presented in Fig. \ref{edgeconcurrence}(a). Near the frequency of topological edge state ($\Delta\sim\pm110\gamma$), strong coherent interactions can be achieved while maintaining a relatively small spontaneous emission rate as dissipative channels, as also clearly implied by the eigenwavefunction in Fig. \ref{edgeconcurrence}(b). By letting the detuning of the QEs be $\Delta/\gamma=-110$, we have $\Gamma_{11}/\gamma_I=\Gamma_{22}/\gamma_I=0.0046$, $J_{12}/\gamma_I=-4.1203$, $\Gamma_{12}/\gamma_I=-6.225\times10^{-4}$.
A density matrix treatment is implemented to investigate the evolution of the two-QE system in a photonic environment modified by the cold atomic metasurface in the Born-Markovian regime: $		\partial_t\rho=i[\rho,H_\mathrm{QE}]/\hbar+\sum_{i,j=1,2}\Gamma_{ij}(2\sigma_{i}\rho\sigma_{j}^\dag-\sigma_i^\dag\sigma_j\rho-\rho\sigma_i^\dag\sigma_j)/2$, where $\rho$ is the $4\times4$ density matrix of two QEs under the normal basis $\left\{|00\rangle, |01\rangle, |10\rangle,|11\rangle \right\}$ \cite{wangPRApp2020}. The Hamiltonian $H_\mathrm{QE}$ for two identical QEs is given by
$H_\mathrm{QE}=\sum_{i=1,2}\hbar \omega_I\sigma_i^\dag\sigma_i+J_{12}(\sigma_1^\dag\sigma_2+\sigma_2^\dag\sigma_1)$, where the environment-induced Lamb shift is neglected without any effect on the result \cite{gonzaleztudelaPRL2011}. The matrix element $\rho_{22(33)}$ represents the population when the first (second) QE is in the ground state and the second (first) QE is in the excited state.  The obtained time evolution of density matrix elements is presented in Fig. \ref{edgeconcurrence}(c), an oscillatory behavior implying a strong exchange of excitations between two QEs is observed, and the decay of population is significantly suppressed compared to the free-space situation. We use two-qubit concurrence, defined as $C(\rho)=\max\left\{0,\lambda_1-\lambda_2-\lambda_3-\lambda_4\right\}$, to measure the entanglement between two QEs \cite{woottersPRL1998}, where $\lambda_i$-s are eigenvalues of the matrix $\sqrt{\sqrt{\rho}\tilde{\rho}\sqrt{\rho}}$ in
decreasing order and $\tilde{\rho}$ is the spin-flipped state, which is defined as $\tilde{\rho}=(\sigma_y\otimes\sigma_y)\rho^*(\sigma_y\otimes\sigma_y)$ with $\rho^*$ the complex conjugate of $\rho$. 
The time evolution of concurrence is given in Fig. \ref{edgeconcurrence}(d), which indicates the entanglement between two QEs can reach a significant value near 0.8 and persist for a very long time. Therefore entanglement is generated without the need for the QEs to be individually addressable, because topological edge states can mediate this entanglement. 
We note the phenomenon that topological edge state can mediate strong coherent interactions between impurity QEs and in the meantime lead to a strong suppression of their decay is a unique property of HOTI systems. This can be seen from the eigenstate wavefunction given in Fig. \ref{edgeconcurrence}(b). For edge states in conventional (first order) topological systems, the rise of coherent interactions also makes the decay of impurity QEs increase, leading to the short lifetime of entanglement. 
Due to the robustness against disorder, the present platform enables resilient generation of entanglement and permits the generation of entangled quantum network for a large number of QEs if we construct a quantum network using these topological edge states. 


In summary, this work provides a thorough understanding of high-order quantum optics in cold atomic settings.
Further explorations include non-Markovian effects involving the corner and edge states, non-Hermitian effects and experimental realizations as well as analogs in other systems like nitrogen-vacancy centers. The topologically protected, strong photon-atom interaction enables the generation of robust and long-lived quantum entanglement in a scalable way without the need of additional photonic structures, promising for quantum information sciences. Additionally, the intrinsic nonlinearity of photon-atom interaction can further facilitate the exploration of the interplay between high-order topology and many-body quantum physics \cite{perczelPRL2020,zhongPRL2020}.

This work has been financially supported by the National Natural Science Foundation of China (Grants No. 51906144 and No. 51636004) and Shanghai Key Fundamental Research Grant (Grants No. 18JC1413300 and No. 20JC1414800). 

\bibliography{ssh2d}
\end{document}


\clearpage
	\title{Supplemental Material for "High-order topological quantum optics in ultracold atomic metasurfaces"}
\author{B. X. Wang}
\affiliation{Institute of Engineering Thermophysics, School of Mechanical Engineering, Shanghai Jiao Tong University, Shanghai 200240, China}
\author{C. Y. Zhao}
\email{changying.zhao@sjtu.edu.cn}
\affiliation{Institute of Engineering Thermophysics, School of Mechanical Engineering, Shanghai Jiao Tong University, Shanghai 200240, China}

\date{\today}
\maketitle
\onecolumngrid
\setcounter{equation}{0}
\setcounter{figure}{0}
\setcounter{section}{0}
\renewcommand{\thesection}{S.\Roman{section}}
\renewcommand{\theequation}{S\arabic{equation}}
\renewcommand{\thefigure}{S\arabic{figure}}
\renewcommand{\citenumfont}[1]{S#1}
\renewcommand{\bibnumfmt}[1]{[S#1]}
\section{Band structure calculation of infinite lattices}
\subsection{General formalism}
%

The two-level atom is assumed to have three degenerate excited states denoted by $|e_{\alpha}\rangle$ polarized along different directions, where $\alpha=x,y,z$ stands for Cartesian coordinates, with a ground state denoted by $|g\rangle$. By applying the single excitation approximation \cite{kaiserJMO2011,kaiserFP2012,guerinPRL2016}, we can work in the subspace spanned by the ground states $|G\rangle\equiv|g...g\rangle$ and the single excited states $|i\rangle\equiv|g...e_i...g\rangle$ of the atoms \cite{kaiserJMO2011,kaiserFP2012,guerinPRL2016}. By adiabatically eliminating the photonic degrees of freedom, the effective Hamiltonian in the absence of the external field is given by \cite{antezzaPRL2009,antezzaPRA2009,kaiserJMO2011,kaiserFP2012,guerinPRL2016,yelinPRL2017,asenjogarciaPRX2017,asenjogarciaPRA2017} 
\begin{equation}\label{Hamiltonian}
	\begin{split}
		H=\hbar\sum_{j=1}^N\sum_{\alpha=x,y,z}(\omega_0-i\frac{\gamma}{2})\ket{e_{j,\alpha}}\bra{e_{j,\beta}}+\frac{3\pi\hbar\gamma c}{\omega_0}\sum_{i=1,i\neq j}\sum_{\alpha,\beta=x,y,z}G_{0,\alpha\beta}(\mathbf{r}_j,\mathbf{r}_i)\ket{e_{j,\alpha}}\bra{e_{i,\beta}},
	\end{split}
\end{equation}
where $\hbar$ is the Planck's constant, $\omega_0$ is angular frequency of the dipole transition of a single atom in free space with a radiative linewidth of $\gamma$, and $c$ is the speed of light in vacuum. The Green's tensor is given by \cite{yelinPRL20172,markelPRB2007,paulusPRE2000}
\begin{equation}
\begin{split}
G_{0,\alpha\beta}(\mathbf{r}_j,\mathbf{r}_i)=-\frac{\exp{(ikr)}}{4\pi r}\Big[\Big(1+\frac{i}{kr}-\frac{1}{(kr)^2}\Big)\delta_{\alpha\beta}+\Big(-1-\frac{3i}{kr}+\frac{3}{(kr)^2}\Big)\hat{r}_{\alpha}\hat{r}_{\beta}\Big]
\end{split}
\end{equation}
In the following we consider out-of-plane modes in which all atoms are excited to the $|e_z\rangle$ states, assuming the in-plane modes can be shifted to other frequencies:
\begin{equation}\label{Hamiltonian2}
	\begin{split}
		H=\hbar\sum_{i=1}^N(\omega_0-i\frac{\gamma}{2})\ket{e_{i,z}}\bra{e_{i,z}}+\frac{3\pi\hbar\gamma c}{\omega_0}\sum_{i=1,i\neq j}G_{0,zz}(\mathbf{r}_j,\mathbf{r}_i)\ket{e_{j,z}}\bra{e_{i,z}},
	\end{split}
\end{equation}

We consider out-of-plane modes in which the polarization of the dipoles are vertical to the 2D plane, namely, only $p_z\neq0$. This amounts to retaining the $G_{0,zz}$ component
\begin{equation}
	G_{0,zz}(\mathbf{r})=\frac{\exp{(ikr)}}{4\pi r}\left(\frac{i}{kr}-\frac{1}{k^2r^2}+1\right).
\end{equation}

For convenience, we define the position of the center of unit cell $(m,n)$ as $\mathbf{R}_{mn}=m\mathbf{a}_1+n\mathbf{a}_2$ with $\mathbf{a}_1=a_0[1~0~0]^T$ and $\mathbf{a}_1=a_0[0 ~1~0]^T$, and the positions of four atoms inside a unit cell are given by $\mathbf{s}_{ij}$ with $i,j=1,2,..,4$, as presented in Fig. 1 in the main text. By invoking the Bloch theorem, for an infinite periodic lattice, we construct the eigenstate wavefunction with an in-plane wavevector $\mathbf{k}=k_x\hat{x}+k_y\hat{y}$ as a linear combination of the single-excited states as \cite{kimPRB2020}
\begin{equation}
|\psi_{\mathbf{k}}\rangle=\sum_{m,n=-\infty}^{\infty}\exp(i\mathbf{k} \cdot\mathbf{R}_{mn})[p_{A,\mathbf{k}}\ket{e_{mnA,z}}+p_{B,\mathbf{k}}\ket{e_{mnB,z}}+p_{C,\mathbf{k}}\ket{e_{mnC,z}}+p_{D,\mathbf{k}}\ket{e_{mnD,z}}],
\end{equation}
where $mn$ denotes the $(m,n)$-th unit cell, $|e_{mn\sigma,z}\rangle$ stand for the single excited states of the $\sigma$-type atom with $\sigma=A,B,C,D$, and $p_{\sigma,\mathbf{k}}$ denotes corresponding expansion coefficients depending on $\mathbf{k}$, and $\mathbf{R}_{mn}=m\mathbf{a}_1+n\mathbf{a}_2$ is the position vector of the center of $(m,n)$-th unit cell. 
We can solve non-Hermitian eigenstate problem 
\begin{equation}
H|\psi_\mathbf{k}\rangle=\hbar E_{\mathbf{k}}\ket{\psi_\mathbf{k}}
\end{equation}
and obtain
\begin{equation}
(\omega_0-i\frac{\gamma}{2}-E_{\mathbf{k}})\sum_{m,n=-\infty}^{\infty}\sum_{\sigma=1}^{4}\exp(i\mathbf{k} \cdot\mathbf{R}_{mn})p_{\sigma,\mathbf{k}}\ket{e_{mn\sigma,z}}+\frac{3\pi\gamma c}{\omega_0}\sum_{m,n=-\infty}^{\infty}\sum_{p,q=-\infty}^{\infty}\sum_{\sigma=1}^{4}\sum_{\tau=1}^{4}G_{zz}(\mathbf{R}_{mn}+\mathbf{s}_\sigma,\mathbf{R}_{pq}+\mathbf{s}_\tau)p_{\tau,\mathbf{k}}\exp(i\mathbf{k} \cdot\mathbf{R}_{pq})\ket{e_{mn\sigma,z}}=0,
\end{equation}
with $\mathbf{R}_{pq}+\mathbf{s}_i\neq\mathbf{R}_{mn}+\mathbf{s}_j$. Equivalently according to the biorthogonality of each singly excited state, we have
\begin{equation}
(-\omega_0+i\frac{\gamma}{2}+E_{\mathbf{k}})\exp(i\mathbf{k} \cdot\mathbf{R}_{mn})p_{\sigma,\mathbf{k}}=\frac{3\pi\gamma c}{\omega_0}\sum_{p,q=-\infty}^{\infty}\sum_{\tau=1}^{4}G_{zz}(\mathbf{R}_{mn}+\mathbf{s}_\sigma,\mathbf{R}_{pq}+\mathbf{s}_\tau)p_{\tau,\mathbf{k}}\exp(i\mathbf{k} \cdot\mathbf{R}_{pq}).
\end{equation}
In a more compact form we have
\begin{equation}
\begin{split}
(-\omega_0+i\frac{\gamma}{2}+E_{\mathbf{k}})p_{\sigma,\mathbf{k}}=-\frac{3\pi\gamma c}{\omega_0}\sum_{\tau=1}^4\mathcal{H}_{\sigma\tau}(\mathbf{k})p_{\tau,\mathbf{k}}
\end{split}
\end{equation}
with elements of the effective Hamiltonian $\mathcal{H}$ in $\mathbf{k}$-space given by
\begin{equation}
	\mathcal{H}_{\sigma\tau}(\mathbf{k})=-\sum_{m,n=-\infty}^{\infty}G_{zz}(\mathbf{R}_{mn}+\mathbf{s}_\sigma,\mathbf{s}_\tau)\exp{(i\mathbf{k}\cdot\mathbf{R}_{mn})}.
\end{equation}
More specifically, if $\sigma=\tau$
\begin{equation}
	\mathcal{H}_{\sigma\sigma}(\mathbf{k})=-\sum_{m,n=-\infty, \mathbf{R}_{mn}\neq0}^{\infty}G_{zz}(\mathbf{R}_{mn},0)\exp{(i\mathbf{k}\cdot\mathbf{R}_{mn})}.
\end{equation}
and if $\sigma\neq \tau$
\begin{equation}
	\mathcal{H}_{\sigma\tau}(\mathbf{k})=-\sum_{m,n=-\infty}^{\infty}G_{zz}(\mathbf{R}_{mn}+\mathbf{s}_{\sigma\tau},0)\exp{(i\mathbf{k}\cdot\mathbf{R}_{mn})}.
\end{equation}
with $\mathbf{s}_{\sigma\tau}=\mathbf{s}_{\sigma}-\mathbf{s}_\tau$.

Therefore, the complex eigenfrequency (energy) $E_\mathbf{k}$ of the Bloch (quasi)eigenstate is obtained, which can be described by $E_\mathbf{k}=\omega_\mathbf{k}-i\Gamma_\mathbf{k}/2$ with $\omega_\mathbf{k}$ denoting the angular frequency and $\Gamma_\mathbf{k}$ the radiative linewidth of the eigenstate. More specifically, the eigenfrequencies are given by
\begin{equation}
	\begin{split}
		\left[\frac{\omega_{\mathbf{k}}-i\Gamma_{\mathbf{k}}/2}{\gamma}-\frac{\omega_0 - i \gamma/2}{\gamma}\right]p_{\sigma,\mathbf{k}}=-\frac{3\pi }{k_0}\Big[\sum_{\tau=1}^4\mathcal{H}_{ji}(\mathbf{k})p_{\tau,\mathbf{k}}\Big]
	\end{split}
\end{equation}
with $k_0=\omega_0/c$.  We further denote $\Delta=\mathrm{Re}E_{\mathbf{k}}-\omega_0$ as the detuning of the eigenstate, and $\Gamma=-2\mathrm{Im}E_{\mathbf{k}}$ as the corresponding radiative linewidth. Therefore we have

\begin{equation}
	\begin{split}
		\frac{\Delta}{\gamma}=-\frac{3\pi }{k_0}\mathrm{Re}\{\mathrm{Eig}[\mathcal{H}]\}
	\end{split}
\end{equation}
and
\begin{equation}
	\begin{split}
		\frac{\Gamma}{\gamma}=1+\frac{6\pi }{k_0}\mathrm{Im}\{\mathrm{Eig}[\mathcal{H}]\},
	\end{split}
\end{equation}
where $\mathrm{Eig}[\cdot]$ denotes the eigenvalues of a matrix. Note here the matrix elements of effective Hamiltonian are not unique, depending on the choice of unit cell \cite{atalaNaturephys2013}. Applying the periodic gauge the unit cell is chosen such that the matrix elements fulfills $\mathcal{H}_{ij}(\mathbf{k})=\mathcal{H}_{ij}(\mathbf{k}+\mathbf{K})$ with $\mathbf{K}$ denoting a reciprocal lattice vector \cite{lingOE2015,ozawa2018topological}.

\subsection{Calculation of diagonal terms of $\mathcal{H}$}
Here we use the technique developed by Simovski to evaluate the diagonal terms of the effective Hamiltonian in $\mathbf{k}$-space \cite{simovskiJEWA1999}. The diagonal terms $\mathcal{H}_{\sigma\sigma}(\mathbf{k})$ are given by:
\begin{equation}\label{diagonal_sum}
\begin{split}
\mathcal{H}_{\sigma\sigma}(\mathbf{k})&=-\sum_{m,n=-\infty, \mathbf{R}_{mn}\neq0}^{\infty}G_{0,zz}(\mathbf{R}_{mn},0)\exp{(i\mathbf{k}\cdot\mathbf{R}_{mn})}\\
&=\sum_{m,n=-\infty, \mathbf{R}_{mn}\neq0}^{\infty}\frac{\exp{(ikR_{mn})}}{4\pi R_{mn}}\left(\frac{i}{kR_{mn}}-\frac{1}{k^2R_{mn}^2}+1\right)\exp{(i\mathbf{k}\cdot\mathbf{R}_{mn})}
\end{split}
\end{equation}
We specifically consider the slowest-decaying term 
\begin{equation}
\begin{split}
S(\mathbf{k})=\sum_{m,n=-\infty, \mathbf{R}_{mn}\neq0}^{\infty}\frac{\exp{(ikR_{mn})}}{R_{mn}}\exp{(i\mathbf{k}\cdot\mathbf{R}_{mn})},
\end{split}
\end{equation}
which can be evaluated by taking the limit of a more general function as
\begin{equation}
S(\mathbf{k})=\lim_{r\rightarrow0}Q(\mathbf{r},\mathbf{k})-\lim_{z\rightarrow0^+}\frac{\exp{(ikz)}}{z}
\end{equation}
with
\begin{equation}
\begin{split}
Q(\mathbf{r},\mathbf{k})=\sum_{m,n=-\infty}^{\infty}\frac{\exp{(ik|\mathbf{r}-\mathbf{R}_{mn}|)}}{|\mathbf{r}-\mathbf{R}_{mn}|}\exp{(i\mathbf{k}\cdot\mathbf{R}_{mn})}.
\end{split}
\end{equation}
Note, different from $S(\mathbf{k})$, the series $Q(\mathbf{r},\mathbf{k})$ includes the term with $\mathbf{R}_{mn}=0$ and as a result we should subtract this term from the result after taking the limit.

From Poisson's summation formula \cite{tsang2004scattering2}, we have
\begin{equation}
\begin{split}
Q(\mathbf{r},\mathbf{k})=\sum_{mn}\frac{1}{\Omega}F(\mathbf{k}+\mathbf{q}_{mn})\exp{[i(\mathbf{k}+\mathbf{q}_{mn})\cdot\bm{\rho}]}.
\end{split}
\end{equation}
where $\Omega$ is the area of unit cell of the real lattice, and $\mathbf{q}_{mn}$ denotes a reciprocal lattice vector, and $F(\mathbf{p})$ is the (2D) Fourier transform of the function $\exp{(ikr)}/r$ as follows \cite{tsang2004scattering2}
\begin{equation}
F(\mathbf{p})=\int d\bm{\rho}\exp{(-i\mathbf{p}\cdot\mathbf{r})}\frac{\exp{(ikr)}}{r}=\frac{2\pi i\exp{(ik_zz)}}{k_z}
\end{equation}
with $k_z=\sqrt{k^2-p^2}$.
Here $\mathbf{r}=\bm{\rho}+z\hat{z}$ are 3D vectors, while $\bm{\rho}$, $\mathbf{R}$, $\mathbf{p}$, $\mathbf{q}_{mn}$ and $\mathbf{k}$ are 2D vectors.

Combining above equations, we have
\begin{equation}
\begin{split}
S(\mathbf{k})&=\lim_{r\rightarrow0}\sum_{mn}\frac{2\pi i}{\Omega}\frac{\exp{(ik_{z,mn}z)}}{k_{z,mn}}\exp{[i(\mathbf{k}+\mathbf{q}_{mn})\cdot\bm{\rho}]}-\lim_{z\rightarrow0^+}\frac{\exp{(ikz)}}{z}\\
&=\lim_{z\rightarrow0^+}\sum_{mn}\frac{2\pi i}{\Omega}\frac{\exp{(ik_{z,mn}z)}}{k_{z,mn}}{k_{z,mn}}-\frac{\exp{(ikz)}}{z}
\end{split}
\end{equation}
with $k_{z,mn}=k_z(\mathbf{k},\mathbf{q}_{mn})=\sqrt{k^2-|\mathbf{k}+\mathbf{q}_{mn}|^2}$.

For $\mathbf{k}=0$ and taking the limit of $k\rightarrow0_+$, we define the real part of $S(\mathbf{k})$ by (and subtracting the diverging term $R_{mn}=0$ )
\begin{equation}
\begin{split}
D=\lim_{k\rightarrow0_+}\mathrm{Re}S(\mathbf{k}=0)=\lim_{k\rightarrow0_+}\mathrm{Re}\left(\sum_{R_{mn}\neq0}\frac{\exp{(ikR_{mn})}}{R_{mn}}\right)=\lim_{z\rightarrow0_+}\sum_{q_{mn}\neq0}\frac{2\pi }{\Omega}\frac{\exp{(-q_{mn}z)}}{q_{mn}}-\frac{1}{z}.
\end{split}
\end{equation}

Then by taking this limit into account in the general expression, we can get
\begin{equation}\label{LR_summation}
\begin{split}
S(\mathbf{k})&=D+\lim_{z\rightarrow0^+}\sum_{mn}\frac{2\pi i}{\Omega}\frac{\exp{(ik_{z,mn}z)}}{k_{z,mn}}-\frac{\exp{(ikz)}}{z}-\sum_{q_{mn}\neq0}\frac{2\pi }{\Omega}\frac{\exp{(-q_{mn}z)}}{q_{mn}}+\frac{1}{z}
\\&=D+\frac{2\pi i}{\Omega}\frac{1}{\sqrt{k^2-\mathbf{k}^2}}+\lim_{z\rightarrow0^+}\left(\frac{1}{z}-\frac{\exp{(ikz)}}{z}\right)+\frac{2\pi }{\Omega}\lim_{z\rightarrow0^+}\sum_{q_{mn}\neq0}\frac{\exp{[-\sqrt{|\mathbf{k}+\mathbf{q}_{mn}|^2-k^2}z]}}{\sqrt{|\mathbf{k}+\mathbf{q}_{mn}|^2-k^2}}-\frac{\exp{(-q_{mn}z)}}{q_{mn}}\\&=D+\frac{2\pi i}{\Omega}\frac{1}{\sqrt{k^2-\mathbf{k}^2}}-ik+\frac{2\pi }{\Omega}\lim_{z\rightarrow0^+}\sum_{q_{mn}\neq0}\left(\frac{1}{\sqrt{|\mathbf{k}+\mathbf{q}_{mn}|^2-k^2}}-\frac{1}{q_{mn}}\right)
\end{split}
\end{equation}

As discussed by Zhen \textit{et al} \cite{zhenPRB2008}, the interesting quantity $D$ is found to possess definite physical meaning, which represents a geometrical effect in electrostatic limit as $k$ tends to zero, and only depends on the lattice structure rather than any wave nature of the summation. On the other hand, both the second term and third term in above equation depend on the wave number $k$ and wave vector $\mathbf{k}$ so that they mainly represent the wave nature of the summation. Last, the series of correction terms is evaluated over reciprocal lattice excluding origin so that it depends on both the geometrical effect and the wave nature. However, the correction series is much smaller than other terms. As a result, the lattice sum above equation is dominated by the first three terms and is very fast converging. To evaluate $D$, we first replace the discrete summation by a continuous summation, namely
\begin{equation}
\frac{4\pi^2}{\Omega}\sum_{q_{mn}\neq0}\rightarrow\int_0^{2\pi}d\theta\int_{g_\mathrm{min}}^\infty gdg.
\end{equation}
This replacement is assumed to be rigorous as long as the integral is taken outside a finite circle centered at the origin and with radius $g_\mathrm{min}$.
\begin{equation}
\begin{split}
D=&\lim_{z\rightarrow0_+}\sum_{q_{mn}\neq0}\frac{2\pi }{\Omega}\frac{\exp{(-q_{mn}z)}}{q_{mn}}-\frac{1}{z}\\
&=\lim_{z\rightarrow0_+}\left(\frac{1}{2\pi}\int_0^{2\pi}d\theta\int_{g_\mathrm{min}}^\infty gdg\frac{e^{-zg}}{g}-\frac{1}{z}\right)\\&=\lim_{z\rightarrow0_+}\left(\int_{g_\mathrm{min}}^\infty dge^{-zg}-\frac{1}{z}\right)\\&=
=\lim_{z\rightarrow0_+}\left(\frac{e^{-zg_\mathrm{min}}}{z}-\frac{1}{z}\right)\\&=-g_\mathrm{min}
\end{split}
\end{equation}

Then we should determine numerically the value of $g_\mathrm{min}$. For doing this, we introduce a parameter sufficiently large and the let the summation in the above equation be numerically calculated in the range of $0<q_{mn}\leq U/z$. We thus introduce the following quantity
\begin{equation}
\begin{split}
L=&\lim_{z\rightarrow0_+}\sum_{0<q_{mn}\leq U/z}\frac{2\pi }{\Omega}\frac{\exp{(-q_{mn}z)}}{q_{mn}}\\
&=\lim_{z\rightarrow0_+}\left(\int_{g_\mathrm{min}}^{U/z} dge^{-zg}\right)\\&=
=\lim_{z\rightarrow0_+}\left(\frac{e^{-zg_\mathrm{min}}-e^{-U}}{z}\right)
\end{split}
\end{equation}
Thus by numerically calculating $L$, we can get $D=-g_\mathrm{min}$ as
\begin{equation}
D=-g_\mathrm{min}=\lim_{z\rightarrow0_+}\left(\frac{\ln(Lz+e^{-U})}{z}\right)
\end{equation}
Another form of above equation to reach $D$ from numerically calculated $L$ is
\begin{equation}
D=\lim_{z\rightarrow0_+}\left(L+\frac{e^{-U}-1}{z}\right)
\end{equation}
For square lattice $D=-3.9002/d$ \cite{zhenPRB2008,simovskiJEWA1999}. In Ref. \cite{zhenPRB2008}, the short-range and intermediate-range terms of
dipolar Green's function in Eq. (\ref{diagonal_sum}) are summed over $3000\times3000$ grids, while the long-range terms are evaluated by Eq. (\ref{LR_summation}) , with the correction terms summed in the region $|q_{mn}|\leq320/d$.

\subsection{Calculation of off-diagonal terms of $\mathcal{H}$}
Here to analytically carry out the above summation, we use the Ewald summation technique \cite{proctorACSPhoton2019,bettlesPRA2017,tsang2004scattering2,zhenPRB2008}. For summing the long-range term on an infinite and complete 2D lattice, the Ewald’s method can be applied to split the term into two parts, with one fast converging on the spatial lattice and the other on reciprocal lattice \cite{tsang2004scattering2}. For off-diagonal terms, we have
\begin{equation}\label{offdiagonal_sum}
	\begin{split}
		\mathcal{H}_{\sigma\tau}(\mathbf{k})&=\sum_{mn}^{N\rightarrow\infty}G_{zz}(\mathbf{R}_{mn}+\mathbf{s}_{\sigma\tau},0)\exp{(i\mathbf{k}\cdot\mathbf{R}_{mn})}\\&
		=\sum_{mn}^{N\rightarrow\infty}\frac{\exp{(ik|\mathbf{R}_{mn}-\mathbf{s}_{\tau\sigma}|)}}{4\pi |\mathbf{R}_{mn}-\mathbf{s}_{\tau\sigma}|}\left(\frac{i}{k|\mathbf{R}_{mn}-\mathbf{s}_{\tau\sigma}|}-\frac{1}{k^2|\mathbf{R}_{mn}-\mathbf{s}_{\tau\sigma}|^2}+1\right)\exp{(i\mathbf{k}\cdot\mathbf{R}_{mn})}\\
		&=\sum_{mn}^{N\rightarrow\infty}\frac{\exp{(ik|\mathbf{s}_{\tau\sigma}-\mathbf{R}_{mn}|)}}{4\pi |\mathbf{s}_{\tau\sigma}-\mathbf{R}_{mn}|}\left(\frac{i}{k|\mathbf{s}_{\tau\sigma}-\mathbf{R}_{mn}|}-\frac{1}{k^2|\mathbf{s}_{\tau\sigma}-\mathbf{R}_{mn}|^2}+1\right)\exp{(i\mathbf{k}\cdot\mathbf{R}_{mn})}
	\end{split}
\end{equation}
We again consider the following series
\begin{equation}
	\begin{split}
		Q(\mathbf{r},\mathbf{k})=\sum_{mn}^{N}\frac{\exp{(ik|\mathbf{r}-\mathbf{R}_{mn}|)}}{|\mathbf{r}-\mathbf{R}_{mn}|}\exp{(i\mathbf{k}\cdot\mathbf{R}_{mn})}.
	\end{split}
\end{equation}
and we can get
\begin{equation}
	\begin{split}
		Q(\mathbf{r},\mathbf{k})=\sum_{mn}\frac{2\pi i}{\Omega}\frac{\exp{[i\sqrt{k^2-|\mathbf{k}+\mathbf{q}_{mn}|^2}z]}}{\sqrt{k^2-|\mathbf{k}+\mathbf{q}_{mn}|^2}}\exp{[i(\mathbf{k}+\mathbf{q}_{mn})\cdot\bm{\rho}]}.
	\end{split}
\end{equation}
in the reciprocal space. As indicated in Ref. \cite{tsang2004scattering2}, this series in reciprocal lattice converge faster than the direct real domain summation, however, it still does not converge well in some cases with large lattice constants and in-plane evaluating points ($z=0$). In this circumstance, it is better to apply the Ewald's summation method.
To this end, we can split $Q(\mathbf{r},\mathbf{k})$ into two components
\begin{equation}
	Q_1(\mathbf{r},\mathbf{k})=\sum_{mn}\exp{(i\mathbf{k}\cdot\mathbf{R}_{mn})}\frac{2}{\sqrt{\pi}}\int_0^Eds\exp{\left[-|\mathbf{r}-\mathbf{R}_{mn}|^2s^2+\frac{k^2}{4s^2}\right]}	
\end{equation}
\begin{equation}
	Q_2(\mathbf{r},\mathbf{k})=\sum_{mn}\exp{(i\mathbf{k}\cdot\mathbf{R}_{mn})}\frac{2}{\sqrt{\pi}}\int_E^\infty ds\exp{\left[-|\mathbf{r}-\mathbf{R}_{mn}|^2s^2+\frac{k^2}{4s^2}\right]}	
\end{equation}
with the use of the expression of the zeroth order spherical Hankel function of the first kind
\begin{equation}
	h_0^{(1)}=-\frac{ie^{ikr}}{kr}=\frac{2}{i\sqrt{\pi}k}\int_C ds\exp{\left[-r^2s^2+\frac{k^2}{4s^2}\right]}.	
\end{equation}
In the next, we will calculate $Q_1$ in the reciprocal domain and $Q_2$ in the real domain. The final result for $Q_1$ is
\begin{equation}
	\begin{split}
		Q_1(\mathbf{r},\mathbf{k})=\frac{\pi i}{\Omega}\sum_{mn}\frac{\exp{[i(\mathbf{k}+\mathbf{q}_{mn})\cdot\bm{\rho}]}}{k_{z,mn}}\left\{\exp{(ik_{z,mn}z)}\mathrm{erfc}\left(-\frac{ik_{z,mn}}{2E}-Ez\right)+\exp{(-ik_{z,mn}z)}\mathrm{erfc}\left(-\frac{ik_{z,mn}}{2E}+Ez\right)\right\}.
	\end{split}
\end{equation}
with $k_{z,mn}=k_z(\mathbf{k},\mathbf{q}_{mn})=\sqrt{k^2-|\mathbf{k}+\mathbf{q}_{mn}|^2}$ and $\mathrm{erfc}(z)=(2/\sqrt{\pi})\int_z^\infty dwe^{-w^2}$.
And the final result for $Q_2$ is
\begin{equation}
	\begin{split}
		Q_2(\mathbf{r},\mathbf{k})=\sum_{mn}^{N}\frac{\exp{(i\mathbf{k}\cdot\mathbf{R}_{mn})}}{2|\mathbf{r}-\mathbf{R}_{mn}|}\left\{\exp{(ik|\mathbf{r}-\mathbf{R}_{mn}|)}\mathrm{erfc}\left(|\mathbf{r}-\mathbf{R}_{mn}|E+\frac{ik}{2E}\right)+\exp{(-ik|\mathbf{r}-\mathbf{R}_{mn}|)}\mathrm{erfc}\left(|\mathbf{r}-\mathbf{R}_{mn}|E-\frac{ik}{2E}\right)\right\}.
	\end{split}
\end{equation}
The splitting parameter $E$ is optimally chosen such that $Q_1$ and $Q_2$ do not differ by more than several orders of magnitude.
According to Tsang \textit{et al} \cite{tsang2004scattering2}, for square lattice, the splitting parameter can be chosen as
\begin{equation}
	E=\frac{\sqrt{\pi}}{d}
\end{equation}

\section{Calculation of 2D Zak phase}
In this section, we show the calculation of 2D Zak phase. The 2D Zak phase $(\theta_x,\theta_y)$ is calculated as 
\begin{equation}
\theta_j=-(1/2\pi)\int_\mathrm{BZ}d^2\mathbf{k}\mathrm{Tr}[A_j(\mathbf{k})],~j=x,y
\end{equation}
where $(A_j)_{mn}(\mathbf{k})=i\langle\mathbf{u}_{m\mathbf{k}}^{L}|\partial k_j|\mathbf{u}_{n\mathbf{k}}\rangle$ with $|\mathbf{u}_{n\mathbf{k}}\rangle$ denoting the periodic part of the Bloch function $[p_{A,\mathbf{k}}~p_{B,\mathbf{k}}~p_{C,\mathbf{k}}~p_{D,\mathbf{k}}]^\mathrm{T}$ of the $n$-th band  \cite{liuPRL2017,chenPRL2019,xiePRL2019,chenPRB2020}. 
The 2D Zak phase is associated with the bulk polarization $P_j$ in terms of $\theta_j=2\pi P_j$.
To numerically obtain the 2D Zak phase, one can use a Wilson loop \cite{benalcazarPRB2017,benalcazarScience2017}
\begin{equation}
\theta_x=\frac{d}{2\pi}\int dk_yv_x^n(k_y)
\end{equation}
and
\begin{equation}
\theta_y=\frac{d}{2\pi}\int dk_xv_y^n(k_x),
\end{equation}
where $v_j^n$ is the $n$-th eigenvalue of the Wannier Hamiltonian 
\begin{equation}
H_{w,j}(\mathbf{k})=-i\log\Pi_{q=0}^M\left[F_{j,\mathbf{k}+q\Delta k_j}\right]
\end{equation}
for $M$ satisfying $(M+1)\Delta k_j=2\pi/d$ and $j=x,y$, and the $(m,n)$-th component of $(F_{j,\mathbf{k}})_{mn}=\int\mathbf{u}_{m\mathbf{k}}^{L*}(\mathbf{r})\mathbf{u}_{n,\mathbf{k}+\Delta k_j}(\mathbf{r})d\mathbf{r}$ for $m,n\in1,2,3,...N_\mathrm{occ}$, where $N_\mathrm{occ}$ is the number of bands below the band gap \cite{liuPRB2018,xiePRB2018,kimPRB2020}. The corresponding biorthogonal condition of the periodic parts of left and right Bloch wavefunctions is given by \cite{wuPRA2020}
\begin{equation}
\int\mathbf{u}_{m\mathbf{k}}^{L*}(\mathbf{r})\mathbf{u}_{n,\mathbf{k}}(\mathbf{r})d\mathbf{r}=\delta_{mn}.
\end{equation}

\begin{figure}[htbp]
	\centering
	\includegraphics[width=1\linewidth]{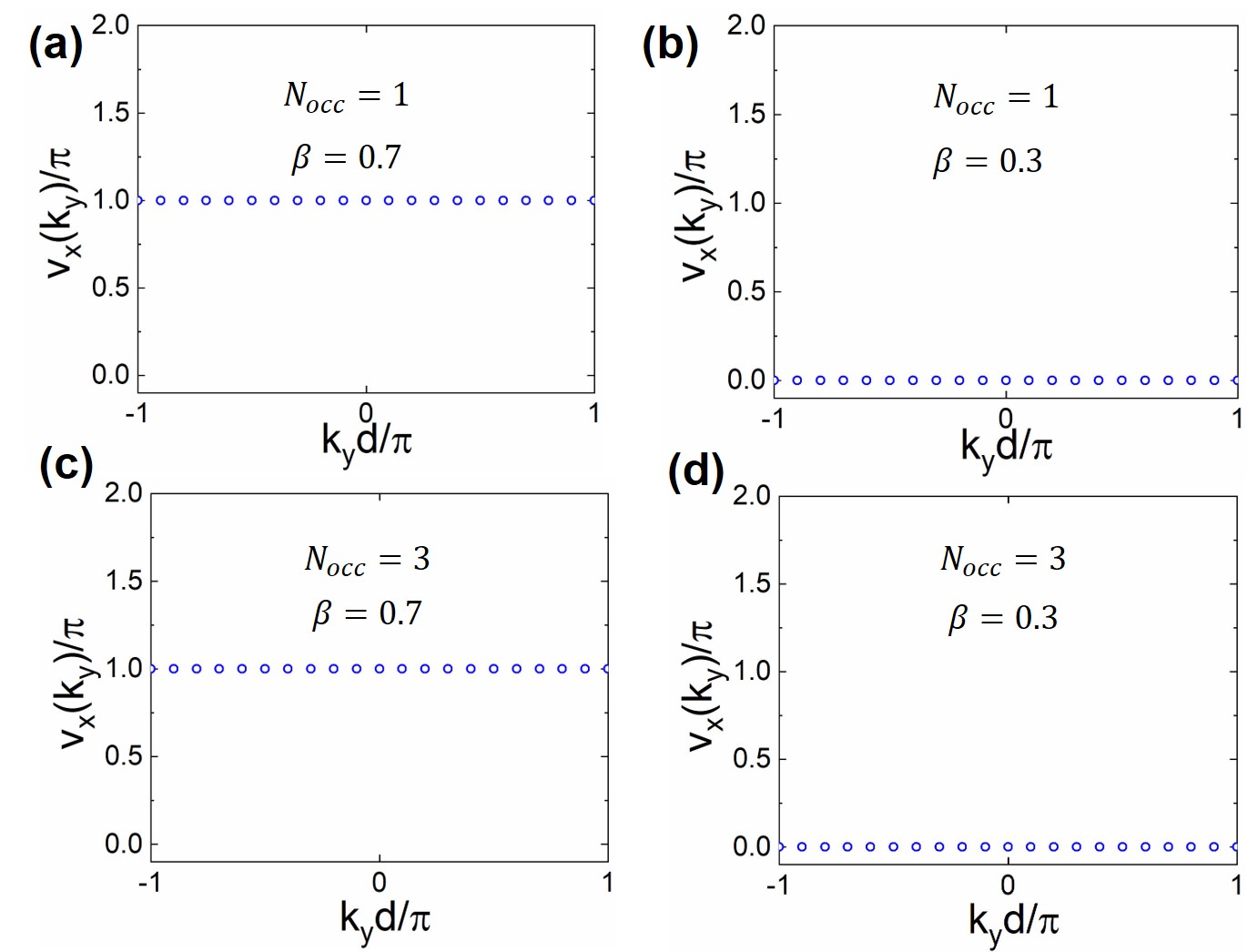}
	\caption{Wannier bands $v_x^n(k_y)$ of the first band gap for the cases under (a) $\beta=0.7$ and (b) $\beta=0.3$, as well as the third band gap for (c) $\beta=0.7$ and (d) $\beta=0.3$.}\label{Wannierbands}
\end{figure}

More specifically, $v_x^n(k_y)$ is the $n$-th eigenvalue of the Wannier Hamiltonian $H_{w,x}(k_y)$ and $v_y^n(k_x)$ is the $n$-th eigenvalue of the Wannier Hamiltonian $H_{w,y}(k_x)$. As an example, in Fig. \ref{Wannierbands}, we show Wannier bands $v_x^n(k_y)$ of the first and third band gaps for topologically nontrivial and trivial cases. It can be clearly found that for the topologically nontrivial cases, $\theta_x$ is $\pi$, which becomes zero in the topologically nontrivial cases. The same is true for $\theta_y$.

\section{Band structure calculation of semi-infinite lattices with domain walls}
In this section, we show the procedure to obtain the band structure of semi-infinite lattices with domain walls of different topology. Without loss of generality, we can assume that the lattice extends infinitely along the $x$-direction with $k_x$ becomes a good quantum number. For a semi-infinite lattice, we can still use the formulas presented in Eqs. (\ref{offdiagonal_sum}) and (\ref{diagonal_sum}), with the difference being that there are a large number of atoms $N_0$ in a single unit cell with domain walls. Here more specifically we consider the domain wall between a topologically nontrivial lattice and free space.

For the out-of-plane polarization, $\mathcal{H}$ is a $N_0\times N_0$ matrix. The diagonal terms are given by
$\mathcal{H}_{jj}(\mathbf{k})$:
\begin{equation}\label{diagonal_sum_interface}
	\begin{split}
		\mathcal{H}_{jj}(\mathbf{k})&=-\sum_{m\neq0}^{N}G_{0,zz}(\omega,k_x|ma_0|,0)\exp{(ik_xma_0)}\\
		&=\sum_{m\neq0}^{N}\frac{\exp{(ik|ma_0|)}}{4\pi |ma_0|}\left(\frac{i}{k|ma_0|}-\frac{1}{k^2m^2a_0^2}+1\right)\exp{(ik_xma_0)}.
	\end{split}
\end{equation}
This is a common summation typically encountered in 1D periodic systems, which can be evaluated as \cite{wang2018topological}
\begin{equation}\label{diagonal_sum_interface2}
	\begin{split}
		\mathcal{H}_{jj}(\mathbf{k})&		=\sum_{m=1}^{\infty}\frac{\exp{(ik|ma_0|)}}{4\pi |ma_0|}\left(\frac{i}{k|ma_0|}-\frac{1}{k^2m^2a_0^2}+1\right)\exp{(ik_xma_0)}+\sum_{m=-\infty}^{-1}\frac{\exp{(ik|ma_0|)}}{4\pi |ma_0|}\left(\frac{i}{k|ma_0|}-\frac{1}{k^2m^2a_0^2}+1\right)\exp{(ik_xma_0)}\\
		&=\sum_{m=1}^{\infty}\frac{\exp{(ikma_0+ik_xma_0)}}{4\pi ma_0}\left(\frac{i}{kma_0}-\frac{1}{k^2m^2a_0^2}+1\right)+\sum_{m=1}^{\infty}\frac{\exp{(ikma_0-ik_xma_0)}}{4\pi ma_0}\left(\frac{i}{kma_0}-\frac{1}{k^2m^2a_0^2}+1\right)\\
	&=\frac{\mathrm{Li}_1(z^+)+\mathrm{Li}_1(z^-)}{4\pi a_0}+i\frac{\mathrm{Li}_2(z^+)+\mathrm{Li}_2(z^-)}{4\pi ka_0^2}-\frac{\mathrm{Li}_3(z^+)+\mathrm{Li}_3(z^-)}{4\pi k^2a_0^3}.
	\end{split}
\end{equation}
with $z^+=\exp{[i(k+k_x)a_0]}$ and $z^-=\exp{[i(k-k_x)a_0]}$ and $\mathrm{Li}_s(z)=\sum_{n=1}^\infty z^n/n^s$ being the polylogarithm.

For off-diagonal terms, we have
\begin{equation}\label{offdiagonal_sum_interface}
	\begin{split}
		\mathcal{H}_{ji}(\mathbf{k})&=-\sum_{m}^{\infty}\mathbf{G}_0(\omega,\mathbf{R}_{m}+\mathbf{s}_{ji},0)\exp{(i\mathbf{k}\cdot\mathbf{R}_{m})}\\&
		=\sum_{m}^{\infty}\frac{\exp{(ik|\mathbf{R}_{m}-\mathbf{s}_{ij}|)}}{4\pi |\mathbf{R}_{m}-\mathbf{s}_{ij}|}\left(\frac{i}{k|\mathbf{R}_{m}-\mathbf{s}_{ij}|}-\frac{1}{k^2|\mathbf{R}_{m}-\mathbf{s}_{ij}|^2}+1\right)\exp{(i\mathbf{k}\cdot\mathbf{R}_{m})}.
	\end{split}
\end{equation}
Note the convergence of above summation is much faster than the complete 2D summation. For convenience, we can directly carry out the summation without resorting to special techniques. We also calculate the IPRs of eigenvectors of the Hamiltonian $\mathcal{H}_{jj}$ for each $k_x$ as an indicator of the degree of localization:
\begin{equation}
	\mathrm{IPR}=\frac{\sum_{j=1}^{N}|p_j|^4}{[\sum_{j=1}^{N}|p_j|^2]^2}.
\end{equation}
This quantity can be utilized to identify edge states.

\section{Dynamics of the cold atomic metasurface}
Under external driving, the effective Hamiltonian in the rotating frame becomes \cite{yelinPRL20172,perczelPRA2017,kaiserFP2012}
\begin{equation}\label{Hamiltonian_drive}
	\begin{split}
\mathcal{H}=\hbar\sum_{j=1}^N\sum_{\alpha=x,y,z}\frac{\Omega_j}{2}(\ket{e_{j,\alpha}}\bra{G}+\ket{G}\bra{e_{j,\alpha}})-\hbar\sum_{j=1}^N\sum_{\alpha=x,y,z}(\Delta+i\frac{\gamma}{2})\ket{e_{j,\alpha}}\bra{e_{j,\beta}}+\frac{3\pi\hbar\gamma c}{\omega_0}\sum_{i=1,i\neq j}\sum_{\alpha,\beta=x,y,z}G_{\alpha\beta}(\mathbf{r}_j,\mathbf{r}_i)\ket{e_{j,\alpha}}\bra{e_{i,\beta}},
	\end{split}
\end{equation}
where $\Omega_j$ is the Rabi frequency of the driving field evaluated at the positions of the atoms given by $\Omega_j=\Omega_0\exp{(i\mathbf{k}\cdot\mathbf{r}_j)}$ in which $\Omega_0=d_{eg} E_L$ with $d_{eg}$ being the transition dipole moment of the atom and $E_L$ the amplitude of the laser field.

Since we only consider the out-of-plane polarization, the wavefunction is assumed in the following form
\begin{equation}\label{tdwavefunction}
	|\psi(t)\rangle=\alpha(t)\ket{G}+\sum_{j=1}^{N}p_{j}(t)\ket{e_{j,z}}.
\end{equation}
Under this effective Hamiltonian and taking the low-excitation limit $\alpha\rightarrow1$, the time-dependent Schr\"odinger equation is 
\begin{equation}
\mathcal{H}|\psi(t)\rangle=i\hbar\frac{d}{dt}|\psi(t)\rangle.
\end{equation}
and then we have 

\begin{equation}
\sum_{j=1}^N\frac{\Omega_j}{2}\ket{e_{j,z}}-\sum_{j=1}^N(\Delta+i\frac{\gamma}{2})p_j(t)\ket{e_{j,z}}+\frac{3\pi\gamma c}{\omega_0}\sum_{i=1,i\neq j}G_{zz}(\mathbf{r}_j,\mathbf{r}_i)p_i(t)\ket{e_{j,z}}=i\frac{d}{dt}\sum_{j=1}^{N}p_{i}(t)\ket{e_{j,z}}.
\end{equation}

In a compact matrix form, the dynamics of light-atom interactions is therefore given by  Ref. \cite{ciprisPRA2021,manzoniNJP2018}
\begin{equation}
\frac{d}{dt}|\mathbf{p}(t)\rangle=\mathbf{M}|\mathbf{p}(t)\rangle+\mathbf{w},
\end{equation}
where 
\begin{equation}\begin{split}
M_{ij}=i(\Delta+i\gamma/2)\delta_{ij}-i\frac{3\pi\gamma }{k}G_{zz}(\mathbf{r}_j,\mathbf{r}_i),
\end{split}
\end{equation}
and $|\mathbf{w}\rangle=-i[\Omega_1\Omega_2...\Omega_j...\Omega_N]/2$.

\begin{figure}
	\centering
	\includegraphics[width=1\linewidth]{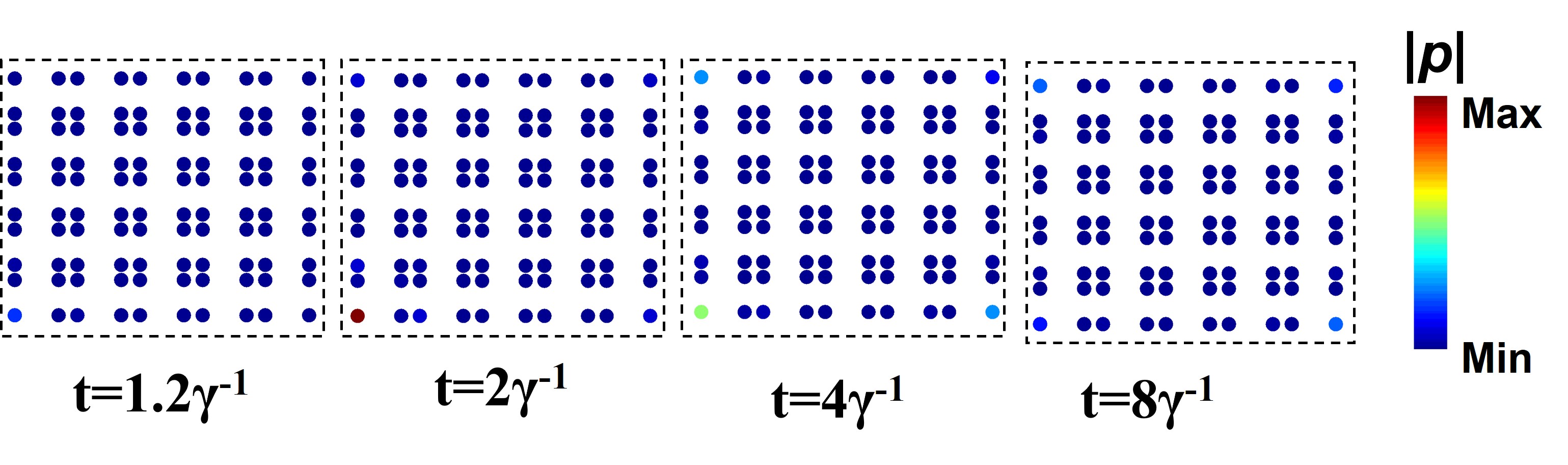}
	
	\caption{Snapshots of the atomic wavefunction at different time moments after the bottom left corner atom is addressed by an external laser drive.}
	\label{cornerevolution_wavefunction}
\end{figure}

In the main text, a Gaussian-shape source field $\Omega_s(t)$ is applied to excite the bottom left corner atom with a tunable frequency detuning $\Delta$. The Rabi frequency of the source field is chosen to be small enough so that the excitation of the system follows mainly on the collective decay from the system rather than the Rabi oscillation \cite{zhangCommPhys2019}. 
Similar to Ref. \cite{yelinPRL20172}, the driving laser in this work is assumed to be adiabatically switched on with a Gaussian profile $\Omega(t)=\Omega\exp{(-[t-1.5\gamma^{-1}]^2/[0.15\gamma^{-2}])}$ for $t<1.5\gamma^{-1}$. It should be noted that sometimes the profile can be switched using a Sigmoid profile to avoid exciting non-resonant modes and it continuously excites the atom. Another approach is to assume a single excited atom is prepared at the bottom left corner. In Fig. \ref{cornerevolution_wavefunction}, the snapshots of the atomic wavefunction at different time moments after the bottom left corner atom is addressed by an external laser drive.

\section{Calculation of Purcell factor}
The decay rate of an impurity quantum emitter near the cold atom metasurface can be obtained by computing the Green's tensor evaluated at the emitter's position in the presence of the metasurface \cite{asenjogarciaPRX2017,asenjogarciaPRA2017,changRMP2018}. For doing so, we start from the effective Hamiltonian with external driving
\begin{equation}\label{Hamiltonian_drive2}
	\begin{split}
		\mathcal{H}=\mathcal{H}_\mathrm{in}-\hbar\sum_{j=1}^N\sum_{\alpha=x,y,z}(\Delta+i\frac{\gamma}{2})\ket{e_{j,\alpha}}\bra{e_{j,\beta}}+\frac{3\pi\hbar\gamma c}{\omega_0}\sum_{i=1,i\neq j}\sum_{\alpha,\beta=x,y,z}G_{\alpha\beta}(\mathbf{r}_j,\mathbf{r}_i)\ket{e_{j,\alpha}}\bra{e_{i,\beta}},
	\end{split}
\end{equation}
where $H_\mathrm{in}$ is associated with the input field that drives the atoms and here we do not specify its detailed form, unlike in Eq. \eqref{Hamiltonian_drive}.
Given the evolution of the atomic state under $\mathcal{H}$, any observables associated with the total field operator $\hat{\mathbf{E}}_\mathrm{out}(\mathbf{r})$ can be derived from the input-output relation \cite{asenjogarciaPRX2017,manzoniNJP2018}
\begin{equation}
	\label{output1}
	\hat{\mathbf{E}}_\mathrm{out}(\mathbf{r}) = \hat{\mathbf{E}}_\mathrm{in}(\mathbf{r}) + \mu_0d_{eg}\omega_0^2\sum_{j} \mathbf{G}_0(\mathbf{r},\mathbf{r}_j)\hat{\mathbf{d}}_j\sigma^{ge}_j,
\end{equation}
where we have noted that $\gamma=\omega^3d_{eg}^2/(3\hbar\varepsilon_0c^3)$, and $d_{eg}$ is the dipole matrix element associated with the transition, and $\hat{\mathbf{d}}_j$ is unit atomic polarization vector. This equation tells us that the total field is a superposition of the incoming field and the fields emitted by the atoms, whose spatial pattern is determined by the Green's function. 
We further assume the incident (driving) field is generated by a dipole source (impurity quantum emitter) $\bm{\mu}$ and can be expressed in the form of expectation value as
\begin{equation}
\mathbf{E}_\mathrm{in}(\mathbf{r})=	\langle\hat{\mathbf{E}}_\mathrm{in}(\mathbf{r})\rangle=\frac{\omega^2}{\varepsilon_0c^2}\mathbf{G}_0(\mathbf{r},\mathbf{r}_s,\omega)\cdot\bm{\mu}
\end{equation}
Under this illumination we can solve the wavefunction in the form of $	|\psi\rangle=\alpha\ket{G}+\sum_{j=1}^{N}p_{j}|e_{j,z}\rangle$ according to the Schr\"odinger equation describing the evolution of wavefunction under the effective Hamiltonian. Then the expectation value of total field for the obtained wavefunction Eq. \eqref{tdwavefunction} is 
\begin{equation}
	\label{output2}
	\mathbf{E}_\mathrm{out}(\mathbf{r}) = \frac{\omega^2}{\varepsilon_0c^2}\mathbf{G}_0(\mathbf{r},\mathbf{r}_s,\omega)\cdot\bm{\mu} + \frac{\omega^2d_{eg}}{\varepsilon_0c^2}\sum_{j} \mathbf{G}_0(\mathbf{r},\mathbf{r}_j,\omega_{eg})\cdot \hat{\mathbf{d}}_j\langle\sigma^{ge}_j\rangle.	
\end{equation}
By noting $\langle\sigma^{ge}_j\rangle=p_j$ with respect to the obtained wavefunction  after assuming $\alpha\approx1$ \cite{kaiserJMO2011}, we get
\begin{equation}
	\label{output3}
	\mathbf{E}_\mathrm{out}(\mathbf{r}) = \frac{\omega^2}{\varepsilon_0c^2}\mathbf{G}_0(\mathbf{r},\mathbf{r}_s,\omega)\cdot\bm{\mu} + \frac{\omega^2d_{eg}}{\varepsilon_0c^2}\sum_{j} \mathbf{G}_0(\mathbf{r},\mathbf{r}_j,\omega_{eg})\cdot \hat{\mathbf{d}}_jp_j.	
\end{equation}

Again, the Green's tensor accounting for the presence of the cold atomic metasurface can be expressed equivalently as \cite{asenjogarciaPRX2017}
\begin{equation}
		\mathbf{E}_\mathrm{out}(\mathbf{r})=\frac{\omega^2}{\varepsilon_0c^2}\mathbf{G}(\mathbf{r},\mathbf{r}_s,\omega)\cdot\bm{\mu}.
\end{equation}
Thus by combing above two equations we can solve the Green's tensor for the vacuum modified by the presence of the cold atomic metasurface given an arbitrary position vector $\mathbf{r}$. Afterwards the decay rate of the impurity quantum emitter in such photonic environment under the Markovian approximation is
\begin{equation}
	\Gamma_I(\mathbf{r}_s,\omega)=\frac{\pi\omega}{3\hbar\varepsilon_0}|\bm{\mu}|^2\rho_{\bm{\mu}}(\mathbf{r}_s,\omega),
\end{equation}
where
\begin{equation}
	\rho_{\bm{\mu}}(\mathbf{r}_s,\omega)=\frac{6\omega}{\pi c^2}\hat{n}_{\bm{\mu}}\cdot\mathrm{Im}\left[\mathbf{G}(\omega,\mathbf{r}_s,\mathbf{r}_s)\right]\cdot\hat{n}_{\bm{\mu}}.
\end{equation}
is the partial local density of states (LDOS) projected to the polarization direction of the impurity quantum emitter in the vacuum modified by the presence of cold atomic arrays, with $\hat{n}_{\bm{\mu}}$ being the unit vector of the dipole moment of the impurity quantum emitter. The decay rate of such emitter in vacuum is
\begin{equation}
	\gamma_I(\mathbf{r}_s,\omega)=\frac{\pi\omega}{3\hbar\varepsilon_0}|\bm{\mu}|^2\rho_0=\frac{\omega^3}{3\hbar\varepsilon_0c^3}|\bm{\mu}|^2.
\end{equation}
where $\rho_0=\omega^2/(\pi^2 c^3)$ is the density of states in vacuum.

Therefore the Purcell factor in the Markovian approximation is obtained from the full Green's function as
\begin{equation}
	F_P=\frac{\Gamma_I}{\gamma_I}=\frac{\rho_{\bm{\mu}}(\mathbf{r}_s,\omega)}{\rho_0}=\frac{6\pi c}{\omega}\hat{n}_{\bm{\mu}}\cdot\mathrm{Im}\left[\mathbf{G}(\omega,\mathbf{r}_s,\mathbf{r}_s)\right]\cdot\hat{n}_{\bm{\mu}}.
\end{equation}
\bibliography{ssh2d}